\documentclass[conference]{IEEEtran}
\IEEEoverridecommandlockouts
\usepackage{cite}
\usepackage{amsmath,amssymb,amsfonts}
\usepackage{algorithmic}
\usepackage{graphicx}
\usepackage{textcomp}
 \usepackage{booktabs}
\usepackage{xcolor}
\def\BibTeX{{\rm B\kern-.05em{\sc i\kern-.025em b}\kern-.08em
    T\kern-.1667em\lower.7ex\hbox{E}\kern-.125emX}}

\begin{document}

\title{Adaptive Hypergraph Network for Trust Prediction\\}

\author{
\IEEEauthorblockN{ Rongwei Xu$^{a}$, Guanfeng Liu$^{a}$, Yan Wang$^{a}$, Xuyun Zhang$^{a}$, Kai Zheng$^{b}$, Xiaofang Zhou$^{c*}$}
\IEEEauthorblockA{\textit{$^a$ School of Computing, Macquarie University, Australia}}
\IEEEauthorblockA{\textit{$^b$ School of Computer Science and Engineering, University of Electronic Science and Technology of China, China}}
\IEEEauthorblockA{\textit{$^c$ Department of Computer Science and Engineering, Hong Kong University of Science and Technology, China}}
\IEEEauthorblockA{\textit{Email: rongwei.xu@hdr.mq.edu.au, guanfeng.liu@mq.edu.au, yan.wang@mq.edu.au}}
\IEEEauthorblockA{\textit{ xuyun.zhang@mq.edu.au, zhengkai@uestc.edu.cn, zxf@cse.ust.hk}}
 }


\newcommand{\ieno}{\textit{i.e.}}
\newcommand{\etcno}{\textit{etc.}}

\maketitle

\begin{abstract}
Trust plays an essential role in an individual's decision-making. Traditional trust prediction models rely on pairwise correlations to infer potential relationships between users. However, in the real world, interactions between users are usually complicated rather than pairwise only. Hypergraphs offer a flexible approach to modeling these complex high-order correlations (not just pairwise connections), since hypergraphs can leverage hyperedeges to link more than two nodes. However, most hypergraph-based methods are generic and cannot be well applied to the trust prediction task. In this paper, we propose an Adaptive Hypergraph Network for Trust Prediction (AHNTP), a novel approach that improves trust prediction accuracy by using higher-order correlations. AHNTP utilizes Motif-based PageRank to capture high-order social influence information. In addition, it constructs hypergroups from both node-level and structure-level attributes to incorporate complex correlation information. Furthermore, AHNTP leverages adaptive hypergraph Graph Convolutional Network (GCN) layers and multilayer perceptrons (MLPs) to generate comprehensive user embeddings, facilitating trust relationship prediction. To enhance model generalization and robustness, we introduce a novel supervised contrastive learning loss for optimization. Extensive experiments demonstrate the superiority of our model over the state-of-the-art approaches in terms of trust prediction accuracy. The source code of this work can be accessed via https://github.com/Sherry-XU1995/AHNTP.
\end{abstract}


\section{Introduction}
Trust is characterized as an inclination to rely on others, and it represents a recognition of the trustworthiness of a person \cite{rotter1971generalized}. Trust can also be described as an individual's willingness to be influenced by the actions of another person \cite{mayer1995integrative}. Trust relationships in social networks have a significant influence on many aspects of people's lives, \ieno, people's decision-making, product recommendations, financial institutions' assessment of customer credit risk, \etcno. In a social network, for instance, if a user wants to purchase a tennis racket but is unsure which brand to choose, recommendations from trusted friends, favorite athletes, or internet celebrities can greatly impact his/her decision. This illustrates the importance of trust in purchase decisions. From a merchant's perspective, understanding the trust relationships between customers can enhance their ability to identify potential customers. However, in real scenarios of social networks, users are usually unknown to each other and it is difficult to observe explicit trust relationships. In this situation, users can still establish an implicit trust relationship based on the characteristics of trust inference \cite{xu2022attention}. For example, users may trust individuals who share similar interests, even if no direct interactions between them. Trust prediction in social networks leverages user behavior data to predict these implicit trust relationships, typically incorporating various features such as historical transaction records, social network connections, and personal profiles\cite{abdullah2017retweet, lin2020guardian, meo2019trust, ghafari2020survey}.

In the literature, trust prediction methods in social networks can be divided into three categories: (1) propagation-based methods, (2) matrix-based methods, and (3) Graph Neural Network (GNN)-based methods. Propagation-based methods are based on trust propagation rules in social networks. Matrix-based methods predict trust connections between non-adjacent users using various factors, including historical user behavior and personal preferences. However, they often neglect the structure information of the network. In contrast to the two categories, GNN-based methods have shown promise in modeling relational data \cite{wu2020comprehensive}, and thus they have been adopted in trust prediction. For example, the DeepTrust model proposed by Wang et al. \cite{wang2019deeptrust} adopts a deep pairwise neural network to capture the latent features of users from their ratings and reviews on trust prediction. However, it only considers the information of directly connected nodes but does not consider the structure of the networks. Liu et al. \cite{liu2019neuralwalk} design a neural network, NeuralWalk, where they consider indirectly connected nodes to solve the multi-hop trust assessment problem. Furthermore, the model proposed by Lin et al. \cite{lin2020guardian} can learn the rule of trust propagation and aggregation in online social networks with better efficiency than NeuralWalk using GCN. The lack of user features and the difficulties in dealing with multi-aspect properties of social trust limit the capability of the model. Furthermore, to address this problem, Jiang et al. \cite{jiang2022gatrust} develop the GATrust model, which adopts both graph attention networks (GAT) \cite{velickovic2017graph} and GCN to learn the various latent factors of both trustor and trustee.

\begin{figure}[t]
  \centering
  \begin{minipage}[b]{0.4\linewidth}
    \includegraphics[width=\linewidth]{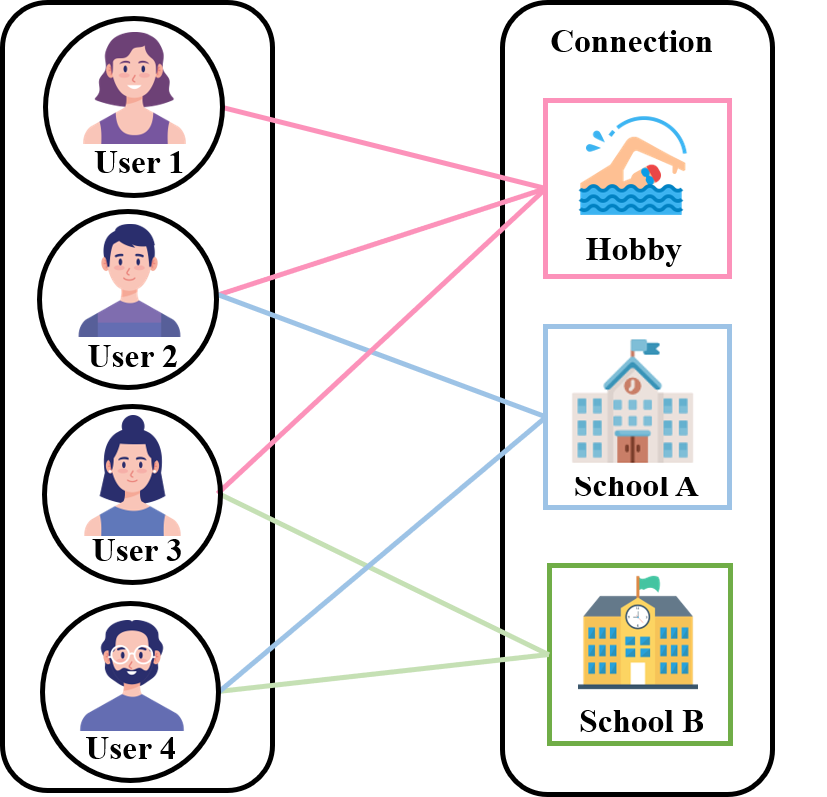}
    \caption{High-order correlation among users and items.}
    \label{fig 2}
  \end{minipage}
  \hfill
  \begin{minipage}[b]{0.5\linewidth}
    \includegraphics[width=\linewidth]{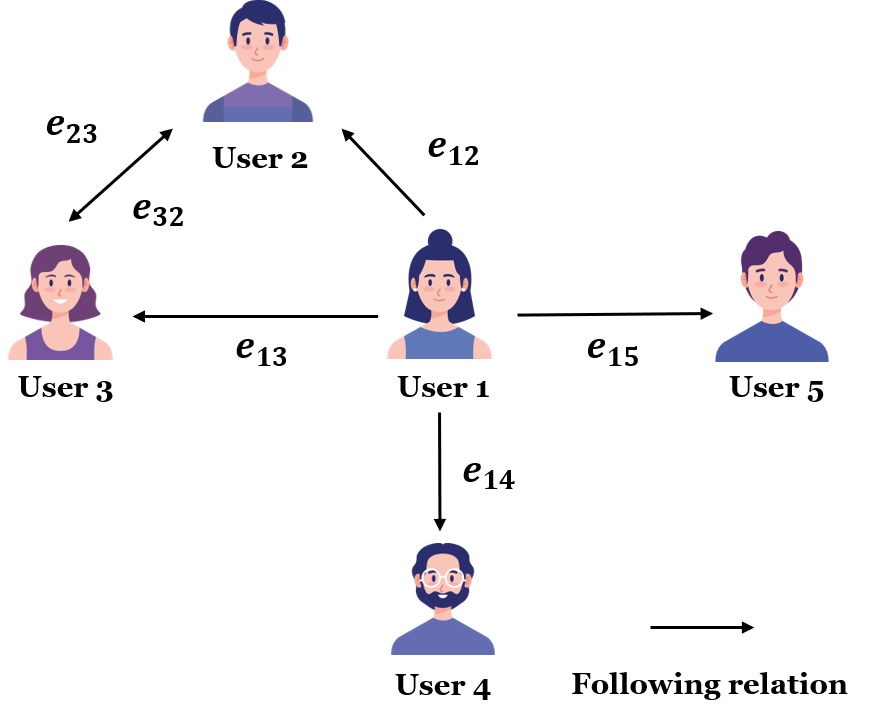}
    \caption{The following relation between five users.}
    \label{fig 1}
  \end{minipage}
\end{figure}

However, most of these GNN-based trust prediction methods cannot directly capture high-order correlations between users. Here, the\textit{ high-order correlation} refers to dependencies that involve more than two nodes rather than simple pairwise correlations \cite{gao2022hgnn}.  For example, in Fig. \ref{fig 2}, Users 1, 2, and 3 share an interest in swimming, and Users 2 and 4 are alumni of the same school. In addition to the pairwise correlations between user-hobby and user-school, there also exist high-order correlations: Users 1, 2, and 3, based on their shared hobby, and Users 2 and 4, according to their school affiliation. These high-order correlations among users are in the pattern of the group, reflecting the information from the social community or social circle. As indicated in social science \cite{cui2018trust}, users have the tendency to maintain a high level of trust in such a specific circle. Incorporating these high-order correlations can produce a more comprehensive embedding, resulting in more accurate trust prediction outcomes. However, representing these high-order correlations effectively in a traditional graph is a challenge.

Firstly, traditional GNN-based methods face difficulties when dealing with isolated nodes. These methods extract user features from pairwise interactions among user-user pairs. They can only capture long-range dependencies in relations (high-order correlation), such as the transitivity of friendship, by aggregating information from neighboring nodes through multiple layers of neural networks \cite{yu2021self}. In the context of online social networks, a prevalence of isolated nodes is observed \cite{guan2023sparse}. Since GNN-based methods heavily depend on pairwise correlation, these isolated nodes can hardly be updated during the learning process. Therefore, they cannot effectively and accurately embed this critical information, which subsequently affects the performance of trust prediction \cite{guan2023sparse}. 

Secondly, GNN-based models face challenges in capturing triangular high-order user relation patterns. For example, as shown in Fig. \ref{fig 1}, the triangular connection among Users 1, 2, and 3 introduces a high-order correlation. These high-order triangle correlations are prevalent and can often convey crucial information in the trust prediction of social networks. Many existing methods primarily focus on first-order pairwise correlations in analyzing the influence of individuals. As shown in Fig. \ref{fig 1}, there are five users, and the edge $e_{ij}$ refers to the directed connection between Users $i$ and $j$. $e_{13}$ indicates User 1 follows User 3. In a traditional graph, the same value is assigned to the original weight of $e_{15}$ and $e_{13}$. However, as observed in Fig. \ref{fig 1}, User 1 has a closer relationship with User 3 than User 5. This is because User 1 follows both User 2 and User 3, while User 2 and User 3 are friends as well. 
\begin{figure}[t]  
	\begin{center}
	\includegraphics[width=2.0in]{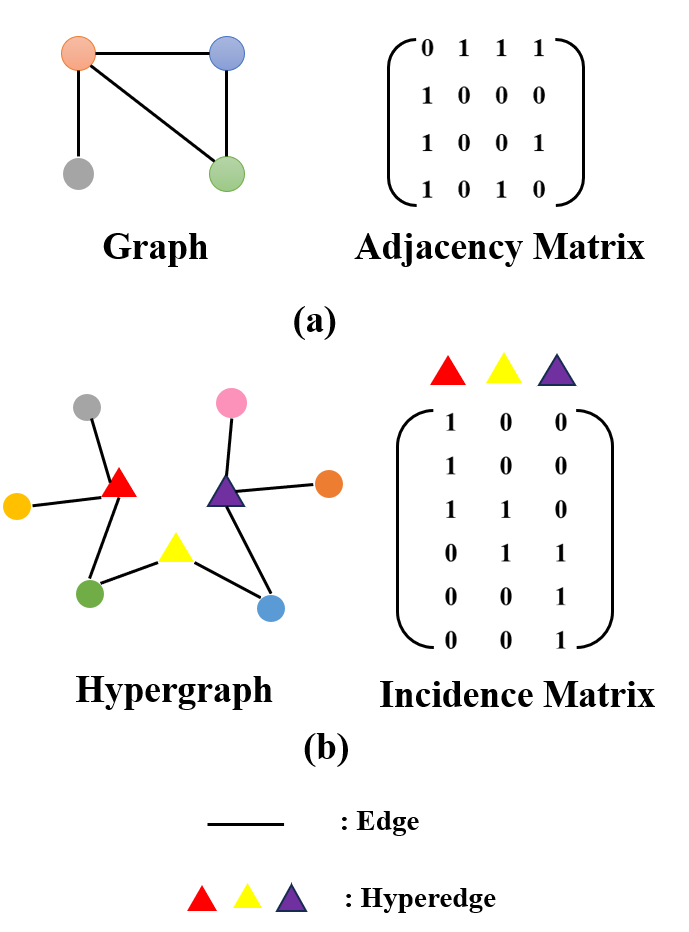}
	\end{center}
	\caption{The comparison between graph and hypergraph.}
	\label{fig 13}
\end{figure}Therefore, we should not set the equal weight for $e_{15}$ and $e_{13}$, since Users 1 and 3 co-occur within this triangular structure. Instead, the weight assigned to $e_{13}$ should be higher than that of $e_{15}$. Therefore, the high-order triangular structures of social influence should be considered when predicting trust relationships.

Hypergraphs \cite{bretto2013hypergraph} can deal with high-order correlations by using hyperedges to connect more than two nodes. This structure can model complex and high-order correlations. As shown in Fig. \ref{fig 13}(a), in a general graph structure, each edge can only connect two nodes. In contrast, as depicted in Fig. \ref{fig 13}(b), in a hypergraph, its hyperedge can link multiple nodes simultaneously, which extends the graph structure to a more comprehensive nonlinear space. A hyperedge can be constructed from various aspects, such as the attribute cluster and the community. High-order correlations can be directly presented on the hyperedge instead of a long message-passing chain in traditional GNN-based methods. Therefore, this approach provides a robust framework for capturing multivariate relationships and high-dimensional data. However, current hypergraphs \cite{bretto2013hypergraph, gao2022hgnn, feng2019hypergraph} are not specifically tailored for trust prediction tasks. Specifically, in the context of social trust prediction, these models face the following three limitations.


 \textit{The first limitation} is that they do not fully consider the social influence of users. The significant role of social influence in trust prediction has been reported in \cite{leskovec2013status, caverlee2008socialtrust, wang2015modeling, tang2016transfer}. The opinions and behaviors of individuals with high social influence, such as internet celebrities or domain experts, tend to have significant influences on social networks. Their suggestions or opinions are more likely to be accepted or trusted by other people \cite{tang2016transfer}. Therefore, the important influence of these high-social-influence people should be considered in the prediction of trust.

\textit{The second limitation} is to distinguish the different influences of hyperedges on different pairs of users. Most of the hypergraph-based methods \cite{feng2019hypergraph, gao2022hgnn, yadati2019hypergcn} rely on a weight matrix to determine the impact of features on output. However, these weights are typically static and uniformly shared across all user pairs, which means they keep constant regardless of the different concerns of users in trust establishment. This approach limits the ability of the model to capture complex and dynamic interactions between users. In the real world, different users can have different concerns in trust establishment. For example, common hobbies might be the key factor in building trust for some users, while for others, similar educational backgrounds might play a more crucial role. Therefore, it's necessary to dynamically adjust the weights of the influential factors based on the specific situation of each user pair. HyperGAT \cite{bai2021hypergraph} has been proposed to incorporate a graph attention mechanism to identify these different influential factors. However, this method requires transforming hyperedges into a homogenous domain of nodes before training. This process may not be able to accurately capture the distinct user interactions. 

\textit{The third limitation} is the loss function of existing methods. Most of those methods utilize cross-entropy loss as their loss function. However, according to the study in \cite{zhang2018generalized}, the use of cross-entropy loss can affect the robustness when dealing with noisy labels. Specifically, when training data contains errors or noisy labels, cross-entropy loss may be sensitive to these labels, which can lead to the overfitting problem. In addition, cross-entropy loss can also lead to the possibility of poor margins \cite{elsayed2018large}, resulting in less distinct boundaries between classes. This can affect the performance, particularly when dealing with datasets characterized by class imbalances. Hence, it is significant to devise a novel loss function that addresses the limitations associated with cross-entropy loss.

To address the limitations mentioned above, we propose a novel Adaptive Hypergraph Network for Trust Prediction (AHNTP). To address the first limitation, we integrate the social influence factor (e.g.,  the triangular motifs in social networks) into the hypergraph structure to capture the high-order correlation in the social network. To address the second limitation, we employ an adaptive Graph Convolutional Network (GCN) to effectively aggregate user information and dynamically identify social influence factors. To address the third limitation, we introduce the contrastive learning method. Specifically, our model utilizes a Motif-based PageRank algorithm \cite{zhao2018ranking} to capture social influence information. It enables the consideration of high-order relationships within specific triangular interactions present in social interactions. Subsequently, we establish a two-tier hypergroup framework. One is based on node-level attributes and the other is based on structure-level attributes. The hypergroup represents a collection of nodes linked by specific correlations. On the basis of different types of correlation, we establish different hypergroups. This approach not only facilitates capturing the correlations but also effectively handles diverse data types and features. In addition, by adopting an adaptive hypergraph GCN, we can dynamically differentiate the influence of each hyperedge on trust evaluation and obtain an accurate representation of users. Finally, we incorporate contrastive learning to improve the generalization and robustness of the model. Experiments on two real datasets have illustrated that AHNTP outperforms state-of-the-art methods in trust prediction. Our contributions are summarized as follows.

\begin{itemize}
    \item[(1)] Firstly, we design a comprehensive and adaptive hypergraph model AHNTP, which can capture the high-order correlation and handle diverse data types and features in trust prediction. 
    \item[(2)] Secondly, we incorporate motif-based PageRank into hypergraphs, which allows us to identify users who have high social influence. In addition, we introduce a multi-hop hypergroup to consider the property of trust propagation. Therefore, our model can embed accurate information of social relationships. 
    \item[(3)] Thirdly, we utilize an adaptive GCN module to distinguish the dynamic impact of different factors on trust prediction. This allows our model to assign higher priority to relevant factors and disregard redundant ones, thereby improving the learning capability of the model.
    \item[(4)] Fourthly, we employ a supervised contrastive learning method to optimize the training loss, which reduces the distances between users with high similarity and increases the distances for those with low similarity.
    \item[(5)] Extensive experiments have been conducted on real-world datasets. The results have demonstrated the proposed AHNTP can outperform the state-of-the-art methods in terms of trust prediction accuracy.
   
\end{itemize}

\section{Related Work}


\subsection{Trust Prediction Methods}
The studies in trust prediction can be classified into three main categories: propagation-based methods, matrix-based methods, and GNN-based trust prediction methods.

\subsubsection{Propagation-based Methods}
Trust propagation-based methods are based on propagation rules in social networks. However, these methods are very time-consuming due to the large size of social networks. Some studies have used path-searching algorithms for efficient trust propagation. For example, Lin et al. \cite{lin2009smallblue} propose a strategy for trust propagation that involves selecting the shortest path between users. However, the varying trust information can be neglected in this method. Hang et al. \cite{hang2008operators} suggest choosing the path with the highest trust value, but cannot fully consider the influence of social relationships. To improve the model, Liu et al. \cite{liu2011finding} propose the MFPB-HOSTP algorithm, incorporating social relationships and recommendation roles. However, it still faces challenges in calculating accurate trust propagation due to attribute imbalance. Gy{\"o}ngyi et al. \cite{gyongyi2004combating} develop the TrustRank algorithm to distinguish quality pages from spam using human-verified seed pages and PageRank. However, the selection of the seed set and the determination of the seed 
distribution can significantly affect the performance of PageRank \cite{prakash2013fuzzy}.

\subsubsection{Matrix-Based Methods}

Matrix-based methods can predict trust relationships between non-adjacent users with the assistance of historical user behavior, personal preferences, and characteristics. They can factorize a trust matrix into the low-rank representation of users and their correlations to approximate the original trust network \cite{wang2019deeptrust}. Yao et al. \cite{yao2013matri} introduce a matrix factorization-based trust prediction model. This method can take advantage of multiple latent factors, prior knowledge, and trust propagation to improve prediction accuracy. Tang et al. \cite{tang2013social} and Zheng \cite{zheng2014trust} introduce a social regularization term for personal preference vectors and a similarity regularization for trust prediction, respectively. In addition, in the work proposed by Liu et al. \cite{liu2013towards}, they consider both social and non-social contexts while building the trust network. Both Liu et al. \cite{liu2013towards} and Huang et al. \cite{huang2013social} utilize factorization to predict missing values in the trust prediction matrix. Meo et al. \cite{meo2019trust} present a matrix factorization-based pairwise trust prediction technique. This method can project a user into two lower-dimensional vectors: the trustor profile and the trustee profile, based on their behavioral bias. However, these methods can lead to cold start issues. Moreover, most of these methods neglect the information on the network structure.

\subsubsection{GNN-based Trust Prediction Methods}
Graph Neural Network (GNN) has demonstrated its effectiveness in feature extraction and modeling the dependencies among nodes in a graph. In the studies \cite{wang2019deeptrust} and \cite{wang2020atne}, they both employ pairwise neural networks for trust prediction. However, these methods do not consider the information on multi-hop connections (indirect connections) between users. In order to address this issue, Xu et al., \cite{xu2022attention} and Liu et al., \cite{liu2019neuralwalk} introduce a multi-hop mechanism and an attention mechanism to capture information between non-neighboring nodes (nodes without direct connection). However, these methods still need to have extensive matrix calculations, making the computation complex. Lin et al., \cite{lin2020guardian} propose a method based on Graph Convolutional Networks (GCN) for trust prediction, which can capture trust propagation and aggregation rules. However, this method faces challenges in handling the multi-aspect features of social networks. To overcome this limitation, Jiang et al. \cite{jiang2022gatrust} adopt the strengths of Graph Attention Networks (GAT) and GCN for trust assessment. However, most of these models mainly focus on pairwise correlations and neglect higher-order correlations, such as triangle correlations between users, which can significantly impact trust prediction.

\subsection{Hypergraph}

Hypergraphs offer a natural way to model high-order correlations \cite{yu2021self}. Ahmad et al. \cite{ahmad2011trust} employ a hypergraph model to capture the intricate relationships and dependencies between accounts, homes, and characters. They analyze the trust network among players engaged in clandestine behavior in an online game. Meanwhile, Somu et al. \cite{somu2017rough} use hypergraphs to tackle the problem of selecting trustworthy cloud services. Recently, GNN has been combined with hypergraphs to obtain a more accurate embedding. For example, the works in \cite{feng2019hypergraph,yadati2019hypergcn,gao2022hgnn} incorporate GCN into hypergraphs to learn complex data correlations from a spectral and spatial perspective. To \begin{figure}[ht]  
	\begin{center}
	\includegraphics[width=2.8in]{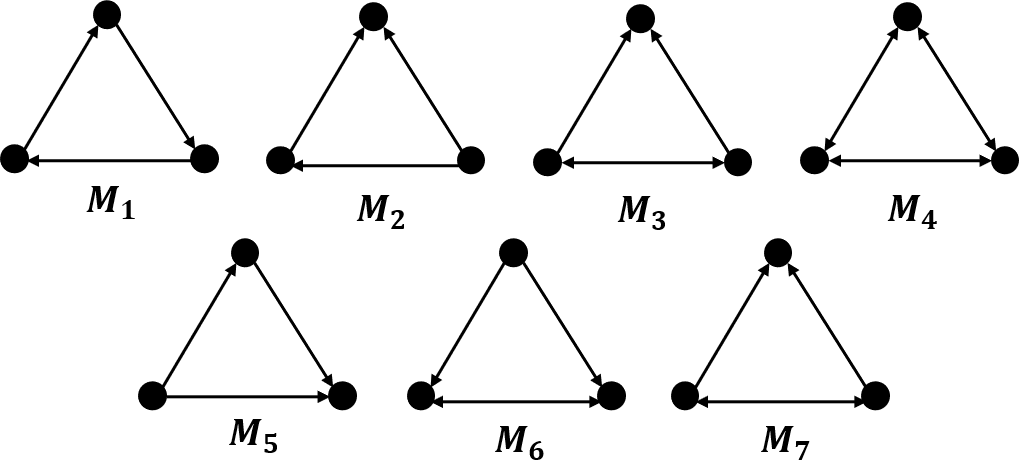}
	\end{center}
	\caption{Seven classical triangular motifs in social networks.}
	\label{fig4}
\end{figure}explore the relationships between different hyperedges, the works in \cite{bai2021hypergraph, xu2022adaptive, ijcai21-UniGNN} adopt an attention mechanism, leading to more discriminative node embeddings. In addition, to integrate the information of structures and attributes, Hu et al, \cite{hu2021adaptive} develop an auto-encoder-based method to learn node embeddings in the hypergraph. However, these methods are too generic and neglect the important social influence factor (e.g., the triangular motifs in social networks).


\section{Preliminary}
\subsection{Hypergraph Definition}
A hypergraph can be presented as $G = (V, E, W)$, where $V$ is the vertex set and $E$ is the hyperedge set. $W$ is the weight matrix of the hyperedge set. $H \in {\left\{0,1\right\}}^{n\times m}$ is an incidence matrix that depicts the node connection in the hypergraph. If the vertex $v$ belongs to the hyperedge $e$, $H_{v,e} = 1$, otherwise $H_{v,e} = 0$. $D_{vv}$ and $D_{ee}$ are presented as the diagonal matrices of the vertex degree and hyperedge degree, respectively.


\subsection{Motif Definition}
A motif network refers to some fundamental mode in a large and complex network \cite{zhao2018ranking}. It consists of a few nodes and edges but has essential influences on the entire network. Studying motif networks can help us understand the high-order structure and function of a network as well as the interplay between nodes. While common motif networks contain triangles, quadrilaterals, and stars, our research focuses solely on the triangular motif because of its frequent occurrence in social networks \cite{zhao2018ranking}. Let $U$ denote a user set. Here, we let $M(R_{U}, {U_{M}})={\left\{(u, S_{U_{M}}(u))|u \in U^{k}\right\}}$ denot a motif set. ${U_{M}} \in {U}$ is an unordered subset of $U$ with $k$ different users and $R_{U}$ is the $k \times k$ adjacency matrix of ${U_{M}}$. As shown in Fig. \ref{fig4}, we still adopt the seven-classical triangular motifs in \cite{zhao2018ranking} to study the high-order relationship between users in social networks. To identify each motif, we define $S_{U_{M}}(u)$ as our selection function, where $u$ represents the order vector of the user.

\subsection{Problem Definition}
A social network can be modeled as a directed hypergraph $G = (U, E, W)$. $U= \left\{u_1, u_2, ..., u_n \right\}$ is denoted as the user set that contains $n$ users. ${R_U} \in {\left\{0,1\right\}}^{n\times n}$ denotes the user-user interaction matrix. We use $H \in {\left\{0,1\right\}}^{n\times m}$ to model the group relationships between users. We denote $X = \left\{X^0_1, X^0_2, ..., X^0_N\right\}, x^0_i \in R^C$ as the initial feature of each vertex, where $C$ is the dimension of features. Given two unknown users, user $u_i$ and user $u_j$, Our objective is to learn a model $f(u_{i}, u_{j}) \rightarrow y \in [0, 1]$, which computes the likelihood of whether $u_i$ and $u_j$ trust each other based on their attributes (e.g., similar hobbies and social community affiliations). The inputs and outputs for this work are as follows:
\begin{itemize}
    \item \textbf{Input:} the user feature matrix $X$, the hyperedge incidence matrix $H$, and the weight matrix of hyperedge $W$.
    \item \textbf{Output:} the probability associated with each class classification, indicating whether a pair of users trust each other or not.
\end{itemize} 

\begin{table}[t]
\footnotesize
\caption{THE NOTATIONS AND EXPLANATIONS IN THIS PAPER}

\begin{tabular}{p{1.5cm}p{6cm}}

\hline

Notation &  Explanation \\

\hline
$G$ & A hypergraph $G = (U, E, W)$. \\
$U$ & A user set. \\
$E$ & A hyperedge set in $G$.  \\
$W$ & A hyperedge set weight matrix in $G$.  \\
$H$ & An incidence matrix depicting the node connection in the hypergraph.\\
$D_{vv}$, $D_{ee}$ & The diagonal matrices of the vertex degree and the hyperedge degree, respectively. \\
$C$ & The dimension of each vertex feature.\\
${R_U}$ & An user-user interaction matrix. \\
$S_{U_{M}}(u)$ & A selection function of motifs. \\
$M$ & A motif set. \\
${U_{M}}$ & An unordered k-tuple subset of $U$. \\
$R$ & The regularizer of the hypergraph.\\
${R_{M}}^{k \times k}$ & An adjacency matrix of ${U_{M}}$. \\
$G'$ & An unweight directed graph $G' = (U, E', R_U)$.\\
$E'$ & An edge set and ${e'_{i,j}}$ connects users $u_i$ and $u_j$. \\
$T_P$ & A transition probability matrix.\\
$d$ & Damping vector.\\
$s$ & The basic PageRank score.\\
$A^M_{ij}$ & A motif-based adjacency matrix.\\
$UC$ & An adjacency matrice of unidirectional connection in G'.\\
$BC$ & An adjacency matrice of bidirectional connection in G'.\\
$\bigodot$ & The Hadamard product.\\
$s'$ & The motif-based PageRank.\\
$Mess^t_e$ & The message of each hyperedge in the t-th layer.\\
$N_e$ & A vertex inter-neighbour set that belongs to the hyperedge.\\
$N_{u_i}$ & A hyperedge inter-neighbour set that belongs to the vertex $u_i$.\\
$h_e$ & The hyperedge feature.\\
$Mess^{t+1}_{u_i}$ & The message of vertex $u_i$ in the (t+1)-th layer.\\
$\theta$ & The trainable parameter.\\
$\beta$ & A share attention mechanism used to compute attention coefficients.\\
$T_U$ & The user embedding after MLP.\\

\hline
\end{tabular}
\label{labn}
\end{table}

\begin{figure*}[ht]
    \centering
    \includegraphics[width=6.5in]{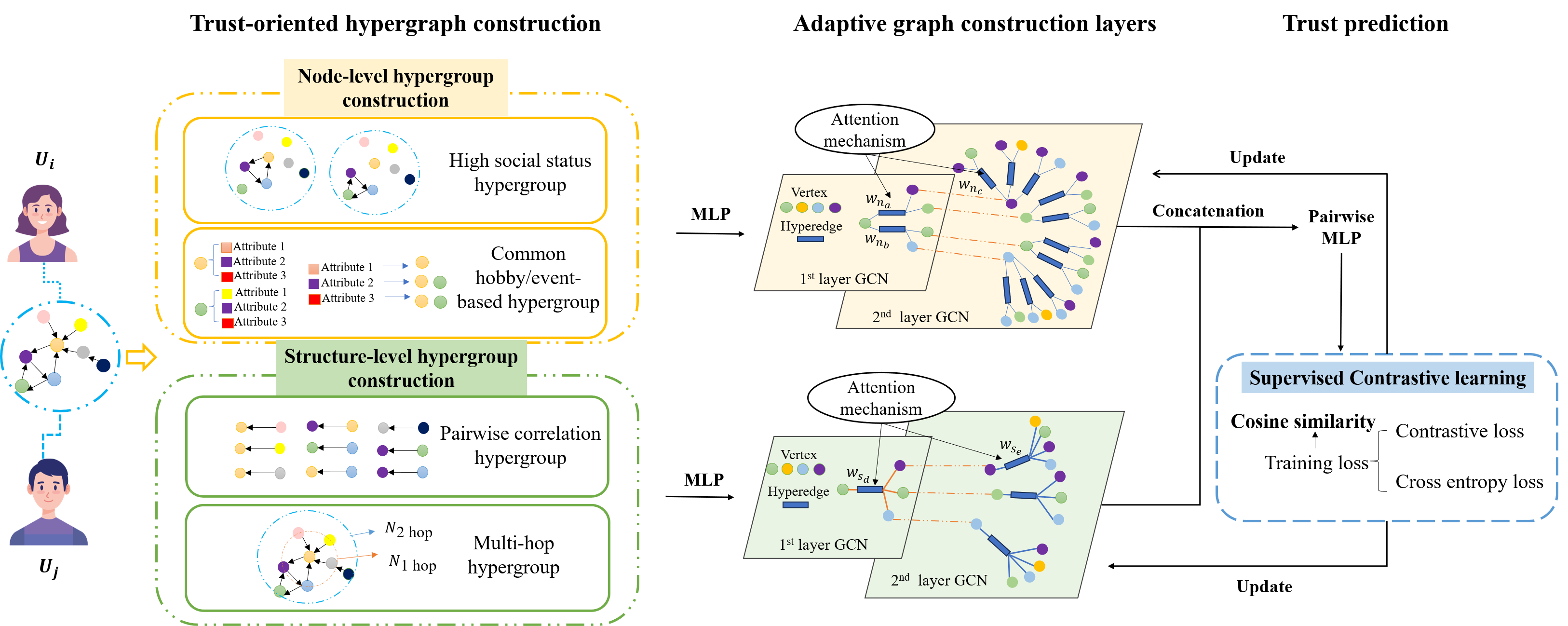}
    \caption{An overview of the proposed AHNTP  model}
    \label{fig:3}
\end{figure*}

\section{The proposed method}
\subsection{Overview}
In this work, we propose an Adaptive Hypergraph Network for Trust Prediction (AHNTP) to accurately predict trust relationships in social networks. As illustrated in Fig. \ref{fig:3}, our approach initially constructs two-level hypergroups: (1) node-level hypergroups, including the social influence hypergroup and the user attribute hypergroup; and (2) structure-level hypergroups, including the pairwise correlation hypergroup and the multi-hop hypergroup. Then, we utilize a Multi-Layer Perceptron (MLP) to extract features from these hypergroups. Afterward, employing the proposed adaptive convolution layer, we aggregate information along the vertex-hyperedge-vertex paths to create the ultimate node embedding. These node embeddings are then concatenated and fed into a pairwise deep neural network to compute their similarities. Finally, we leverage a contrastive learning method to optimize our model. The notations used in this paper are summarized in TABLE \ref{labn}.

\subsection{Trust Hypergraph Construction}
To effectively capture high-order information of both structure and attributes in a social network, we initially built up a trust-oriented hypergraph. This hypergraph integrates node-level attribute-based hypergroups and structure-level attribute-based hypergroups.

\subsubsection{High Social influence Hypergroup}
In this hypergroup, given a user, our goal is to identify the neighbours who have high social influence. Those neighbours can greatly affect people's opinions and decision-making \cite{ford2018we, sokolova2020instagram}. According to the studies in \cite{zhao2018ranking, rachman2013analysis}, a user who interacts with multiple people simultaneously is likely to have high social influence. Consequently, we employ an unsupervised motif-based PageRank algorithm to identify these influential figures.

Let $G' = (U, E', {R_U})$, where $U$ is the set of users. $E'$ is the edge set, where $e' \in E'$. ${e'_{i,j}}$ is an edge that connects users $u_i$ and $u_j$. ${R_U}$ is the weight matrix of $E'$ in an unweighted direct graph. The transition probability matrix can be defined as Eq. (1). The initial basic PageRank score $s$ is calculated by Eq. (2):
\begin{small}
    \begin{align} \label{equ: 1}
    {T_p}=\frac{R_{U_{ij}}}{\sum_{i} {R_{U_{ij}}}}
    \end{align}
\end{small}
\begin{small}
    \begin{align} \label{equ: 2}
    s = d{{T_p}^T}s + {\frac{1-d}{n}}e
    \end{align}
\end{small}where $s_i$ is the original PageRank value of node $i$. $e$ is the unit vector. $d$ is a damping vector at the range of (0,1). $n$ is the number of nodes. 

By Eq. (3), we can derive the motif-based adjacency matrix $A^{M}_{ij}$ which can capture the co-occurrence of two nodes within the specific motif. The process is defined by:
\begin{small}
    \begin{align} \label{equ: 3}
    A^{M}_{ij}= \left\{ 
    \begin{array}{lc}
        \sum 1 & {i,j} \subset S_{U_{M}}, \\
        0& \text{otherwise}. \\
    \end{array}
\right.
    \end{align}
\end{small}

\begin{figure}[ht]  
	\begin{center}
	\includegraphics[width=2.5in]{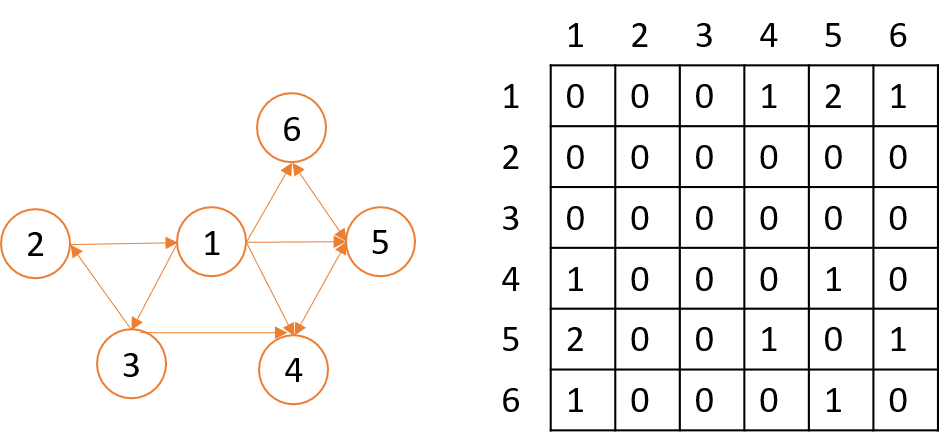}
	\end{center}
	\caption{$M_6$-based adjacency matrix of six nodes.}
	\label{fig5}
\end{figure}

If both $u_i$ and $u_j$ appear in a specific motif, we will add their frequency to the adjacency matrix; otherwise, we add 0 to this matrix. For example, as shown in Fig.\ref{fig5}, $A^{M_6}_{15}$ is 2 since $u_1$ and $u_5$ both appear in $M_6$ (\ieno $\left\{1, 6, 5 \right\}$ and $\left\{1, 5, 4 \right\}$). The greater the value in $ A^{M}_{ij}$, the stronger the influence of $u_j$ towards $u_i$ within the motif.

To obtain the high-order relationship between users in a given motif $M_k$, we calculate the motif-based adjacency matrix $A^{M_k}_{ij}$. We set $UC$ and $BC$ as the adjacency matrices of unidirectional connection and bidirectional connection in $G'$, respectively. Then we have $BC= {R_U}\bigodot{{R_U}^T}$ and $UC={R_U}-BC$, where $\bigodot$ is the Hadamard (entrywise) product. Given two users $u_i$ and $u_j$, where $u_k$ is the intermediate user, the three users in a triangular motif can have six unique position combinations. The co-occurrence frequency of any two users is computed by summing the co-occurrence frequencies of the two users across these six instances. TABLE \ref{lab2} lists the calculation for these seven motif-based adjacency matrices. 

After we obtain the motif-based adjacency matrix, we leverage a linear combination method to combine the high-order structure information (motif-based adjacency matrix) with the edge-based structure information (pairwise-based adjacency matrix). This combination finally forms a comprehensive weight matrix that can be calculated by Eq. (4):
\begin{small}
    \begin{align} \label{equ: 3}
    W_c = \alpha \cdot R_U + (1-\alpha)\cdot A^{M_k} 
    \end{align}
\end{small}where $\alpha \in [0, 1]$ is a parameter to balance the influential factors between the pairwise relation and the high-order relation of users. 

Through a normalization process, the final social influence of the node while considering the high-order correlation information can be calculated by Eq. (5):
\begin{small}
    \begin{align} \label{equ: 2}
    s' = d{\frac{W_{c_{ij}}}{\sum_{i} {W_{c_{ij}}}}}s + {\frac{1-d}{N}}e
    \end{align}
\end{small}

To construct the high social influence hypergroup, we select the top $K$ neighbors with the highest $s'$ scores connected to the given user to form a hyperedge. The concatenation of all these hyperedges forms our high social influence hypergroup, which can be calculated by Eq. (6):
\begin{small}
    \begin{align} \label{equ: 2}
   H_{hss} = H^{1}_{hss} || H^{2}_{hss} || ...|| H^{K}_{hss}
    \end{align}
\end{small}

\begin{table}[t]
\footnotesize
\caption{Computation of $M_1$-$M_7$ motif-induced adjacency matrices}
\scriptsize
\begin{tabular}{p{0.5cm}p{6cm}p{1cm}}

\hline
Motif &  Matrix Computation & $A^{M_k}$\\

\hline
$M_1$ & $C=(UC\cdot UC)\bigodot UC^T$ &  $C+C^T$\\
$M_2$ & $C=(BC \cdot UC) \bigodot UC^T + (UC \cdot BC)\bigodot UC^T + (UC \cdot UC) \bigodot BC$ &  $C+C^T$\\
$M_3$ & $C=(BC \cdot BC) \bigodot UC + (BC \cdot UC) \bigodot BC + (UC \cdot BC) \bigodot BC$ & $C+C^T$ \\
$M_4$ & $C = (BC \cdot BC) \bigodot BC$ & $C$  \\
$M_5$ & $C = (UC \cdot UC) \bigodot UC + (UC \cdot UC^{T} ) \bigodot UC + (UC^T \cdot UC) \bigodot U$  & $C+C^T$\\
$M_6$ & $C = (UC \cdot BC) \bigodot UC + (BC \cdot UC^T ) \bigodot UC^T + (UC^T \cdot UC) \bigodot BC$ & $C$ \\
$M_7$ & $C = (UC^T \cdot BC) \bigodot UC^T + (BC \cdot UC) \bigodot UC + (UC \cdot UC^T ) \bigodot BC$ & $C$ \\
\hline
\end{tabular}
\label{lab2}
\end{table}

\subsubsection{Attribute-based Hypergroup}
In this hypergroup, our goal is to identify users with similar attributes, such as hobbies, ages, living in cities, or participating in the same event. The combination of these different attribute-based hyperedges forms the attribute-based hypergroup, which can be calculated by Eq. (7):
\begin{small}
    \begin{align} \label{equ: 2}
   H_{attr} = H_{attr1} || H_{attr2} || ...|| H_{attrL}
    \end{align}
\end{small}

\subsubsection{Pairwise-based Hypergroup}
In this hypergroup, our objective is to cover the basic pairwise (low-order) correlation between users. Each hyperedge only links two nodes, which transforms a graph structure into a two-uniform group. The synthesis of these pairwise-based hyperedges is combined to create the pairwise-based hypergroup as follows:
\begin{small}
    \begin{align} \label{equ: 2}
   H_{pair} = H_{pair1} || H_{pair2} || ...|| H_{pairM}
    \end{align}
\end{small}

\subsubsection{Multi-hop Hypergroup}
In this hypergroup, our objective is to capture more information by considering multiple hops of neighbors. By leveraging the $N$-hop neighbours, the model can capture more high-order information. The concatenation of these multi-hop hyperedges leads to the construction of the multi-hop hypergroup as follows:
\begin{small}
    \begin{align} \label{equ: 2}
   H_{hop} = H_{hop1} || H_{hop2} || ...|| H_{hopN}
    \end{align}
\end{small}
Before we proceed to the convolution layer, we utilize a Multi-Layer Perceptron (MLP) to extract the latent features of hypergroups. Each hypergroup feature is multiplied by a trainable weight parameter, followed by a transformation that maps the feature to a different dimension space. This process allows us to acquire a comprehensive and accurate feature representation.

\subsection{Adaptive Convolution Layer}
In this section, we will construct a two-step adaptive convolution module derived from the spatial domain. In traditional graph models, a message is transferred directly between the central vertex and its adjacent nodes. In our model, the embedding information of each vertex is transferred via the hyperedge in our model, which can effectively leverage high-order correlation. This message propagation path can be denoted as $P(u_{1}, u_{i}) = (u_1, e_1, u_2, e_2, u_3, ..., u_{i-1}, e_i, u_i)$, where $u_{i-1}$ and $u_i$ are elements of the vertex subset associated with the hyperedge $e_i$. Given a vertex $u_i$, we aim to aggregate the information from the vertex sets that belong to the hyperedges of $u_i$. To achieve this goal, our first step is to aggregate the message of each hyperedge $e$ in the $t$-th layer, which can be calculated by Eq. (10):
\begin{small}
    \begin{align} \label{equ: 2}
    {Mess_{e}^t} = \begin{matrix} \sum_{u_{i} \in N_{e}} \frac{x^{t}_{u_i}}{|N_{e}|} \end{matrix}
    \end{align}
\end{small}where $x^{t}_{u_i}$ is the input feature of $u_i$ in $t$-th layer. $N_{e}$ is the vertex set that belongs to the hyperedge $e$. $\frac{u_{i}}{|N_{e}|}$ is a normalization of vertex feature. 

To update the feature of hyperedges, we employ a trainable weight $w_e$ to multiply with the aggregated information of the hyperedge. The updated hyperedge feature $h_e$ for hyperedge $e$ can be calculated by Eq. (11):
\begin{small}
    \begin{align} \label{equ: 2}
h_e = {w_i} {Mess_{e}^t}
    \end{align}   
\end{small}

Next, our goal is to aggregate the information from the hyperedges that are connected to the vertex $u_i$ in layer $t+1$. The process can be calculated using Eq. (12):
\begin{small}
    \begin{align} \label{equ: 2}
    {Mess_{u_i}^{t+1}} = \begin{matrix} \sum_{e \in N_{u_i}} \frac{h^{t}_{e}}{|N_{u_i}|} \end{matrix}    
    \end{align}   
\end{small}where ${Mess_{u_i}^{t+1}}$ denotes the message of $u_i$ in the $(t+1)$-th layer. $N_{u_i}$ represents the set of hyperedges that are connected to vertex $u_i$. Similarly, $\frac{h^{t}_{e}}{|N_{u_i}|}$ is a normalization of the hyperedge feature. To update the vertex feature in the next layer, we use ReLU as our activation function. This process can be calculated by Eq. (13):
\begin{small}
    \begin{align} \label{equ: 2}
    x^{t+1}_{u_i} = F(x^t_{u_i}, \mathrm{Mess}_{u_i}^{t+1})= \mathrm{Relu}(Mess_{e}^{t+1} \cdot {\theta}^t)    
    \end{align}   
\end{small}where $F()$ is the update function of the vertex and $\theta^t$ denotes the trainable parameter of the $t$-th layer. 

To identify the varying effects of different hyperedges on the central node, we leverage an attention mechanism. Initially, we apply a trainable linear transformation to both the central node and its associated hyperedges. Then we apply a shared attention mechanism, denoted as $\beta$. It can facilitate self-attention on both the central node and the hyperedge. This method can illustrate the significance of the hyperedge on the central vertex. This procedure is calculated by Eq. (14).
\begin{small}
    \begin{align} \label{equ: 2}
    a_{ie} = \sigma(\beta^T [Wx^{t+1}_{u_i}||Wh_e])
    \end{align}   
\end{small}where $W$ is the matrix for weights and $||$ means the concatenation. $\sigma$ is the LeakyReLU activation function. 

Next, we normalize $a_{ie}$ to obtain the final attention coefficient of the indicated hyperedge. This process can be calculated by Eq. (15).

\begin{small}
    \begin{align} \label{equ: 2}
    w_{ie} = \frac{exp(a_{ie})}{\begin{matrix} \sum_{e \in N_{u_i}} exp(a_{ie}) \end{matrix}} = \frac{exp( \sigma(\beta^T [Wx^{t+1}_{u_i}||Wh_e]))}{\begin{matrix} \sum_{e \in N_{u_i}} exp(\sigma(\beta^T [Wx^{t+1}_{u_i}||Wh_e]) \end{matrix}}
    \end{align}   
\end{small}
We reweight the neighboring hyperedges of the central vertex, which can be calculated by Eq. (16).
\begin{small}
    \begin{align} \label{equ: 2}
    x^{t+1}_{u_i} = Relu (\begin{matrix} \sum_{e \in N_{u_i}} w_{ie}Wh_e \end{matrix})
    \end{align}   
\end{small}

To effectively capture the node-level and structure-level features of the trust-oriented hypergraph, we separately process them through the adaptive convolution layer. After generating the embeddings with the adaptive GCN, we concatenate them to form a comprehensive user feature embedding.

Finally, to mine the latent attributes of users, we input their comprehensive embeddings into a pairwise deep neural network. By utilizing this pairwise MLP, we can map user features into a lower-dimensional space. Such mapping can facilitate calculating the similarity between users to conduct trust prediction. This process can be calculated by Eqs. (17) and (18).
\begin{small}
    \begin{align}
    &T_{u_i}=f(...f(W_{a2}f(X_{u_i}^\prime)W_{a1})...)\\
    &T_{u_j}=f(...f(W_{b2}f(X_{u_j}^\prime)W_{b1})...)
\end{align}
\end{small}where $T_{u_i}$ and $T_{u_j}$ are outputs of MLP, $W_a$ and $W_b$ represent weight matrices of the layer. $f()$ is a ReLU activation function, which can enhance the nonlinearity of the model. 

Then we compute the cosine similarity between these two users to evaluate the likelihood of mutual trust as follows,
\begin{small}
    \begin{align}
    CS(T_{u_i},T_{u_j})=\frac{T_{u_i}^T T_{u_j}}{\|T_{u_i}\|\|T_{u_j}\|}
\end{align}
\end{small}where $CS(T_{u_i},T_{u_j})$ is the likelihood that $u_i$ trusts $u_j$. $T_{u_i}$ and $T_{u_j}$ are the final vector representation of $u_i$ and $u_j$ respectively. This output is a continuous cosine similarity score in the range of [0,1], which indicates the degree of trustworthiness between the two users.

\subsection{Supervised Contrastive Learning}

While many existing trust prediction studies utilize cross-entropy loss for model optimization, the recent research \cite{gunel2020supervised} has indicated that cross-entropy can lead to poor generalization performance. To address this problem, we adopt contrastive learning \cite{li2022hmgcl}. In our model, a combined training loss that incorporates contrastive loss and cross-entropy loss is designed to optimize our model. Therefore, in the embedding space, it can keep users with high similarity close to each other, and keep those with low similarity far away from each other. This requires defining both positive and negative samples. We treat user pairs who trust each other in social networks as positive samples, represented as $T_P = {(u_1, u_2),..., (u_i, u_j)}$. In contrast, users who distrust each other are treated as negative samples, denoted as $T_N = {(u_3, u_4),... (u_i, u_k)}$. This contrastive training loss can be calculated by Eq. (20):
\begin{small}
    \begin{align}
   L_1 = -\frac{1}{|U|} {\begin{matrix} \sum_{i \in U} (\log \frac{\begin{matrix} \sum_{(i,j) \in T_P} \exp((CS(T_{u_i},T_{u_j})/t)) \end{matrix}}{\begin{matrix} \sum_{(i,k) \in T_P \cup T_N} \exp((CS(T_{u_i},T_{u_k})/t)) \end{matrix}} \end{matrix}})
\end{align}
\end{small}where $CS(T_{u_i},T_{u_j})$ is the cosine similarity between users. $t$ denotes a temperature parameter, which can adjust the magnitude of the contrastive loss. $L_1$ represents the supervised contrastive training loss, and $U$ is the user node set. 

In addition, the cross-entropy loss as another component of our training loss function can be calculated by Eq. (21):
\begin{small}
\begin{equation}
\begin{split}
     L_2 &= -\frac{1}{|U|} \sum_{(i,j) \in {T_P \cap T_N}} \left(\bar{y}_{ij} \cdot \log(CS(T_{u_i},T_{u_j})\right) \\
     &\quad + \left((1-\bar{y}_{ij}) \cdot \log(1 - CS(T_{u_i},T_{u_j}))\right)
\end{split}
\end{equation}
\end{small}where $\bar{y}_{ij}$ is the relationship ground truth lable of $u_i$ and $u_j$. 

Finally, we utilize a linear combination to incorporate contrastive loss and cross-entropy loss into our whole training loss, which is calculated by Eq. (22):
\begin{small}
    \begin{align}
       L = \lambda_1 \cdot L_1 +\lambda_2 \cdot L_2 
    \end{align}
\end{small}where $L$ is the comprehensive train loss for AHNTP. $\lambda_1$ and $\lambda_2$ are the parameters to control the weight of $L_1$ and $L_2$. 

Fundamentally, our task is to classify the node on the hypergraph. Due to the complex structure and the existence of noise within the network, a regularization technique is adopted in our objective function to make label smoothing, as calculated by Eq. (23):
\begin{small}
    \begin{align}
        \arg\min \left\{L(f) + R(f)\right\}
    \end{align}
\end{small}where $f$ is the classification function. $L(f)$ denotes the training loss, which consists of contrastive loss and cross-entropy loss. $R(f)$ represents the regularizer of the hypergraph, which can be calculated by:
\begin{small}
    \begin{align}
        R(f) = f^T(I - D^{-\frac{1}{2}}_{vv}HWD^{-1}_{ee}H^{T}D^{-\frac{1}{2}}_{vv} )f
    \end{align}
\end{small}

\section{Experiment}
In this section, we conduct experiments on two real-world datasets to evaluate the performance of AHNTP in trust prediction. Our experiments focus on answering the following questions:
\begin{itemize}
    \item \textbf{Q1:} How is the performance of the proposed method AHNTP compared to state-of-the-art models in the accuracy of the trust prediction? \textit{(See Section B)}
    \item \textbf{Q2:} How is the robustness of AHNTP compared with the state-of-the-art models? \textit{(See Section B)}
    \item \textbf{Q3:} How does each component of our model contribute to the performance? \textit{(See Section C)}
    \item \textbf{Q4:} What is the impact of the important hyperparameters on our model? \textit{(See Section D)}
\end{itemize}
\subsection{Experimental Setting}
\subsubsection{Dataset}
To evaluate our trust prediction model, we select two datasets: Epinions \cite{tang2012etrust} and Ciao \cite{tang2012etrust}. Both are from product review sites. In these datasets, users can post reviews and ratings of the products they have purchased. Other users can then view these reviews and ratings, subsequently providing a rating to indicate the level of ``helpfulness" based on the quality of the reviews. Then, after reading the product reviews and ratings, other users can decide whether or not to trust the user who writes the review, which is the ground truth in the experiments. The details of the datasets are listed in Table \ref{lab1}.

\begin{table}[]

\caption{STATISTICS OF DATASETS}
\centering

\centering

\begin{tabular}{p{4cm}p{1.5cm}p{1.5cm}}

\hline

Dataset & Epinions & Ciao \\

\hline

 Number of Users & 8,935 & 4,104 \\
 Number of Items & 21,335 & 75,071 \\
 Number of  Purchase Behaviors & 220,673 & 171,405 \\ 
 Number of Trust Relations & 65,948 & 41,675  \\
 Data sparsity & 0.16523\% & 0.49499\% \\

\hline
\end{tabular}
\label{lab1}
\end{table}

\subsubsection{Baseline}
The baseline methods can be divided into three categories, including traditional network embedding methods, trust prediction methods, and hypergraph-based methods. By comparing our trust prediction model with these three categories of methods, we can provide a comprehensive evaluation of the performance of our model. To keep a controlled environment and fair comparison, we utilize the same input features of all the models. As the network embedding methods and
hypergraph-based methods are directly relevant to our model, we add fully connected layers and the ReLU function to enable them to predict trust. In detail, the baselines can be listed as follows:
\begin{itemize}
    \item[(1)] Traditional network embedding methods:
    \begin{itemize}
        \item GAT \cite{DBLP:conf/iclr/VelickovicCCRLB18}: GAT is a deep learning model for graph-structured data, which employs a self-attention mechanism based on  Graph Neural Networks (GNNs). It can weigh and aggregate features of neighbor nodes, which can effectively capture the relevance between nodes.
    
        \item SGC \cite{DBLP:conf/icml/WuSZFYW19}: SGC is a variant of Graph Convolutional Network (GCN). It simplifies the architecture by removing nonlinear activation functions and collapsing the weight matrices between consecutive layers, which speeds up the training process.
    \end{itemize}
    For this group of baselines, we can investigate the performance of our method in trust prediction in different graph embedding methods.
    \item[(2)] Trust prediction methods:
    \begin{itemize}
            \item Guardian \cite{lin2020guardian}: Guardian incorporates information from the structure of social networks into its GCN model to assess the trustworthiness of relationships.
    
            \item AtNE-Trust \cite{wang2020atne}: AtNE-Trust employs auto-encoders and a feature fusion unit to integrate user attributes and trust network structures, thereby enhancing the prediction of trust relationships.
    
            \item KGTrust \cite{yu2023kgtrust}: KGTtrust combines a discriminative convolutional layer with a knowledge-augmented GNN to assess trust relationships. 
    
    \end{itemize}
    For this group of baselines, we can investigate the accuracy of different models in trust prediction.
    \item[(3)] Hypergraph-based methods:
    \begin{itemize}
            \item UniGCN \cite{ijcai21-UniGNN}: UniGCN extends the Graph Convolutional Network (GCN) model to accommodate hypergraph structures.
    
            \item UniGAT \cite{ijcai21-UniGNN}: UniGAT expands the Graph Attention Network (GAT) model to function with hypergraph structures.
    
            \item HGNN+ \cite{gao2022hgnn}: HGNN+ employs a trainable weight matrix to integrate correlations across various modalities or types, capturing high-order interactions through hypergraph convolution in the spatial domain.
 
    \end{itemize}
    For this group of baselines, we can investigate the effectiveness of our proposed method in handling complex and high-order correlations in trust prediction compared to the use of existing hypergraph structures.
\end{itemize}

\subsubsection{Metrics}
To evaluate the effectiveness of the proposed method, we investigated the accuracy of trust prediction and the F1 score of AHNTP and eight baseline models, which are widely used in the literature to measure the performance of models in trust prediction tasks \cite{yu2023kgtrust, lin2020guardian, wang2019deeptrust, wang2020atne}.

\begin{table*}[ht] 
\renewcommand{\arraystretch}{1.4}
\scriptsize
\centering 
\begin{threeparttable}
\caption{Performance comparisons with different training sets on Two datasets} 
\label{tab:my-table} 
\begin{tabular}{c|c|c|cccccccc|cc} 
\hline
\rowcolor{gray!20}
{Dataset} & {Metric (\%)} & {Training Size} & GAT & SGC & Guardian & AtNE-Trust & KGTrust & UNIGCN & UNIGAT & HGNN+ & \textbf{AHNTP} & Improvement$^1$ \\ \hline
\multirow{8}{*}{Ciao}    & \multirow{4}{*}{Accuracy} & 50\% & 59.76 & 67.40 & 71.27 & 62.24 & 71.72 & 74.89 & \underline{82.56} & 82.16 & \textbf{85.12} & 2.56\% \\ 
                         &                           & 60\% & 61.03 & 68.29 & 71.62 & 62.66 & 72.11 & 82.37 & \underline{82.80} & 82.04 & \textbf{85.44} & 2.64\% \\ 
                         &                           &70\% & 62.17 & 68.39 & 71.90 & 63.52 & 72.34 & 82.44 & \underline{83.15} & 82.23 & \textbf{85.56} & 2.41\% \\
                         &                           & 80\% & 63.01 & 68.81 & 71.94 & 66.58 & 72.36 & 83.10 & \underline{83.64} & 82.28 & \textbf{86.11} & 2.48\% \\ \cline{2-13}
                         & \multirow{4}{*}{F1-Score}  & 50\% & 66.47	& 67.53	& 71.84	& 62.76	& 72.85 & 83.39 & \underline{87.63}	& 87.33 &	\textbf{88.90} & 1.27\%\\

                         &                           & 60\% & 68.08 & 68.58 & 72.28 & 63.03 & 73.11 & \underline{87.69} & 87.64 & 87.34 & \textbf{89.36} & 1.67\% \\ 
                         &                           & 70\% & 70.61 & 68.78 & 72.67 & 65.37 & 73.23 & \underline{87.84} & \underline{87.84} & 87.46 & \textbf{89.59} & 1.75\%\\
                         &                           & 80\% & 70.85 & 69.76 & 73.32 & 69.92 & 74.06 & \underline{88.33} & 88.31 & 88.00 & \textbf{90.11} & 1.78\% \\ \hline
\multirow{8}{*}{Epinions} & \multirow{4}{*}{Accuracy} & 50\% & 61.70 & 77.22 & 80.15 & 71.90 & 80.59 & \underline{86.78} & 86.38 & 86.33 & \textbf{89.21} & 2.43\%\\ 
                         &                           & 60\% & 61.92 & 77.57 & 80.22 & 73.01 & 80.65 & \underline{87.52} & 86.59 & 86.39 & \textbf{89.48} & 1.96\% \\ 
                         &                           & 70\% & 64.76 & 77.82 & 80.31 & 73.40 & 80.96 & \underline{87.95} & 86.41 & 86.16 & \textbf{89.55} & 1.60\% \\
                         &                           & 80\% & 70.79 & 78.17 & 80.55 & 73.59 & 81.14 & \underline{87.96} & 86.24 & 86.37 & \textbf{89.78} & 1.82\% \\ \cline{2-13}
                         & \multirow{4}{*}{F1-Score}  & 50\% & 65.60 & 77.63 & 80.41 & 72.87 & 81.05 & \underline{91.11} & 90.77 & 90.78& \textbf{92.51} & 1.40\%\\
                         &                           & 60\% & 66.64 & 77.63 & 80.51 & 73.74 & 81.11 & \underline{91.53} & 90.96 & 90.79 & \textbf{92.75} & 1.22\%\\ 
                          &                           & 70\% & 72.67 & 78.05 & 80.58 & 73.80 & 81.46 & \underline{91.78} & 90.84 & 90.74 & \textbf{92.79} & 1.01\%\\ 
                          &                           & 80\% & 72.84 & 78.56 & 80.86 & 74.29 & 81.70 & \underline{91.79} & 90.83 & 90.92 & \textbf{92.94} & 1.15\%\\ \hline
\end{tabular}
\begin{tablenotes}
\item[The best results are highlighted in bold, while the second-bestest results are underlined. Improvement$^1$ indicates the improvement of our AHNTP compared to the best performing baseline.] 

\end{tablenotes}
\end{threeparttable}
\end{table*}

\subsubsection{Implementation}
All the models are implemented by using Pytorch 1.12 and a cloud server of an NVIDIA RTX 4090 GPU with 24GB of VRAM. This server is also equipped with a 25-core AMD EPYC 7T83 CPU. For experiments involving multi-hop computations, we shift the workload to the CPU due to the VRAM limitation on the GPU. To calculate high-order social influence, we configure the influence factor $\alpha$ in Motif-based PageRank (MPR) to 0.8 and use the resulting rank scores as embedding weights for users. Our network architecture comprises three hypergraph GCN layers, with dimensions of 256, 128, and 64, respectively. We employ the ReLU function as the activation function across all layers. The temperature parameter $t$ in contrastive learning is set to 0.3. As our Epinions and Ciao datasets only contain positive samples (\ieno, user trust lists), we randomly select two pairs of unconnected users as negative samples for training. To optimize our model, we use the Adam optimizer, setting the initial learning rate to 0.001 and the decay weight to 1e-4. The source code of our work is available at https://github.com/Sherry-XU1995/AHNTP.

\subsection{Performance Comparisons \textbf{(Q1 and Q2)}}
\subsubsection{\textbf{To answer Q1}} We compare the proposed model AHNTP with the traditional network embedding methods, trust prediction methods, and hypergraph-based methods. we set 80\% of the dataset as the training set and 20\% of the dataset as the test dataset The experimental results are listed in TABLE IV. Based on this table, we have the following observations and analyses.
\begin{itemize}
    \item \textbf{Observation 1:} AHNTP outperforms all other state-of-the-art methods. In detail, AHNTP achieves an improvement of 11. 25\% in accuracy in the Ciao data set and 9. 18\% in the Epinions data set on average, respectively. Regarding F1-Scores, AHNTP can have an average improvement of 12. 29\% in the Ciao dataset and 10.22\% on the Epinions dataset on average, respectively.
    
    \item \textbf{Analysis 1:} These results demonstrate the superior capability of AHNTP to process graph data on social networks and make accurate trust predictions. Furthermore, the experimental results illustrate the good performance of AHNTP in handling complex and high-order correlations within social networks.

    \item \textbf{Observation 2:} From the result, we can see that the models (HGNNP, UNIGCN,	UNIGAT,	AHNTP, Guardian, KGTrust) incorporating high-order correlations outperform the models that do not incorporate high-order correlation (GAT, SGC, AtNE-Trust). 
    
    \item \textbf{Analysis 2:} This observation indicates the importance of high-order correlations in trust prediction. Compared to methods that only consider pairwise correlations, approaches that introduce high-order correlations provide a better representation of the latent relationships within social networks.
    
    \item \textbf{Observation 3:} Compared to Guadian and KGTrust, which also consider the high-order user correlations, the hypergraph-based methods can have notable improvements in trust prediction. In detail, the accuracy and F1 scores of hypergraph-based methods can be improved by 11. 63\% and 14. 95\% in the Ciao dataset, and by 6. 74\% and 10. 34\% in the Epinions dataset on average, respectively.
    
    \item \textbf{Analysis 3:} Guardian and KGTrust use the adjacency matrix to represent the connection structure of a graph. However, these graph-based methods cannot directly capture high-order information, such as reachable neighbors of k-order, which are crucial for representation learning in graph structures \cite{gao2022hgnn}. Although Guardian and KGTrust incorporate high-order correlations by stacking multiple GCN layers, their performances are still worse than those of the hypergraph-based methods. Furthermore, this approach of stacking multiple GCN layers can fall into the rigid k-hop neighborhood smoothing \cite{gao2022hgnn}. In contrast, the hypergraph-based methods can model high-order correlations more flexibly. Instead of using the other nodes as intermediate nodes in the traditional graph structure, high-order correlations between users can be represented based on hyperedges.

    \item \textbf{Observation 4:} Compared with the other hypergraph-based baselines, AHNTP still has the best performance. In particular, the accuracy and F1-score of AHNTP are 2.48\% and 1.78\% higher than the best baseline, UniGAT, on the Ciao dataset in trust prediction. On the Epinions dataset, the accuracy and F1-score of AHNTP are 1.82\% and 1.55\% higher than the best baseline, UniGCN.
    
    \item \textbf{Analysis 4:} In addition to high-order correlations, our model considers social influence and different preferences on trust establishment, and also our model adopts contrastive learning. Therefore, our model can construct a more accurate representation of the social network.
\end{itemize}

 \subsubsection{\textbf{To answer Q2}}  To investigate the robustness of AHNTP, we set the ratio of the training set as 50\%, 60\%, 70\%, and 80\%, respectively, and set the ratio of the testing set as 20\%. Our key observations are as follows:
 
 \begin{itemize}
     \item \textbf{Observation:} From TABLE IV, we can see that AHNTP still has the best performance under different sizes of training sets. For example, in the Ciao dataset, compared to the best baseline, AHNTP can improve the accuracy of the trust prediction by 2.56\%, 2.64\%, 2.41\%, and 2.48\% with training sets of 50\%, 60\%, 70\%, and 80\%, respectively. Additionally, with the decrease of the training set, there is a corresponding decrease in the accuracy and F1-score of all models. This decrease is more obvious in traditional embedding methods like GAT and SGC, compared with the decrease in AHNTP. For AHNTP, the largest ratio of the decrease in \begin{table}[ht]
\centering
\caption{Ablation Study of Model Variants}
\label{tab:ablation_study}
\begin{tabular}{m{1.6cm}|m{1.5cm}|m{4.2cm}}
\hline
\centering\arraybackslash \textbf{Model} & \centering\arraybackslash \textbf{Removed Part} & \centering\arraybackslash\textbf{Purpose} \\ \hline
\centering\arraybackslash AHNTP$_{\mathrm{nompr}}$ &  \centering\arraybackslash MPR algorithm & \centering\arraybackslash Investigate the impact of high-order correlations in social influence. \\ \hline
\centering\arraybackslash AHNTP$_{\mathrm{noatt}}$ & \centering\arraybackslash  Adaptive GCN layers & \centering\arraybackslash Investigate the impact of our adaptive GCN layer. \\ \hline
\centering\arraybackslash AHNTP$_{\mathrm{nocon}}$ & \centering\arraybackslash Contrastive learning loss & \centering\arraybackslash Investigate the impact of the supervised contrastive learning loss. \\ \hline
\centering\arraybackslash AHNTP & \centering\arraybackslash None & \centering\arraybackslash Investigate the impact with all components. \\ \hline
\end{tabular}
\end{table}accuracy is only 0.99\%, and 1.21\% for the F1-score on the Ciao dataset, even when the size of the training data is reduced from 80\% to 50\%. Similarly, on the Epinions dataset, the largest ratio of the decrease in accuracy is only 0.57\%, and 0.43\% for the F1-score, when the size of the training data is reduced from 80\% to 50\%.

     \item \textbf{Analysis:} These results show the robustness and effectiveness of AHNTP. This is due to the superior capability of AHNTP in representing high-order correlations. AHNTP is not limited to single connections between two entities. Instead, it utilizes information from multiple neighbors. By directly capturing complex and high-order correlations, AHNTP reduces its reliance on multi-step paths. Therefore, it can minimize redundancy during the propagation of information. This means that even when faced with incomplete or inaccurate label data, AHNTP can still keep essential correlation information.  
     
 \end{itemize}

\begin{figure}[ht]
  \centering
  \begin{minipage}[b]{0.49\linewidth}
    \includegraphics[width=\linewidth]{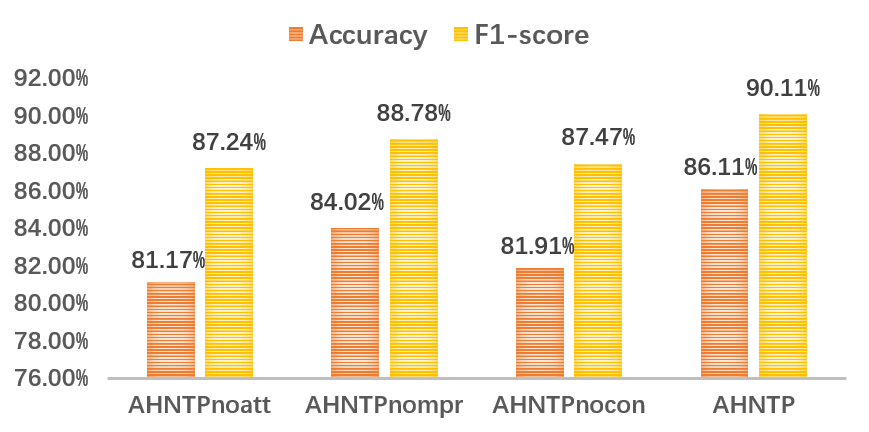}
    \caption{The Performance of Ablation Study on Ciao Dataset.}
    \label{image1}
  \end{minipage}
  \hfill
  \begin{minipage}[b]{0.48\linewidth}
    \includegraphics[width=\linewidth]{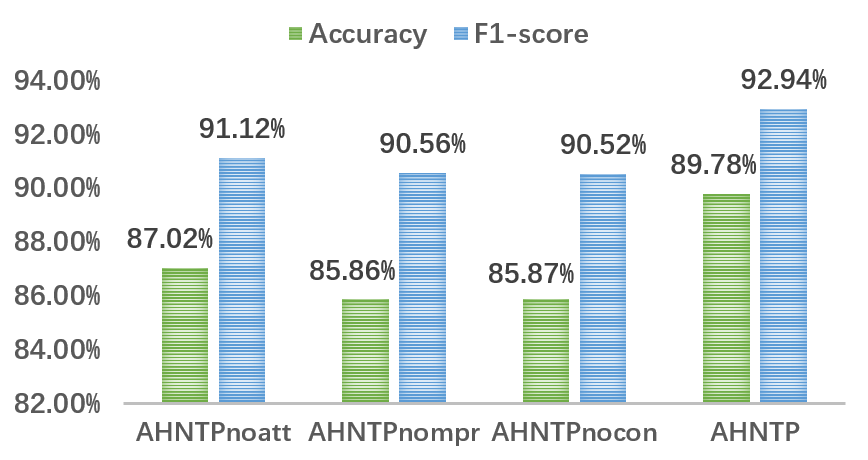}
    \caption{The Performance of Ablation Study on Epinions Dataset.}
    \label{image2}
  \end{minipage}
\end{figure}

\subsection{Ablation Study \textbf{(Q3)}}

\textbf{To answer Q3}, we investigate the impact of each individual component of our model. We compare the following three different variants of AHNTP to verify the effectiveness of each component: 
\begin{itemize}
    \item[(1)] AHNTP$_{\mathrm{nompr}}$: This variant is to evaluate the importance of high-order correlations in social influence for trust prediction, where we discard the MPR algorithm from AHNTP.
    \item[(2)] AHNTP$_{\mathrm{noatt}}$: This variant is to investigate the impact of our adaptive GCN layer, where we replace it with a standard hypergraph convolution layer.
    \item[(3)] AHNTP$_{\mathrm{nocon}}$: This variant is to verify the contribution of the supervised contrastive learning loss, where we replace it with a cross-entropy loss.
\end{itemize}

 Fig. \ref{image1} and Fig. \ref{image2} show the results of the ablation study on the Ciao and Epinion datasets. Based on two figures, we can have the following observations and analysis.

\begin{itemize}
    \item \textbf{Observation:} On the Ciao dataset, AHNTP can reach an accuracy of 86.11\% and an F1-score of 90.11\%. Compared to its variants, AHNTP can improve the accuracy and F1 score of trust prediction by 4.94\% and 2.87\% over AHNTP$_{\mathrm{noatt}}$, 2.09\% and 1.33\% over AHNTP$_{\mathrm{nompr}}$, and 4.20\% and 2.64\% over AHNTP$_{\mathrm{nocon}}$, respectively. In the Epinions dataset, AHNTP can reach an accuracy of 89.78\% and an F1-score of 92.94\%. Here, AHNTP can improve the accuracy and F1 score of trust prediction by 2.76\% and 1.82\% \begin{table}[htbp]
  \centering
  \caption{Multi-hop Experiments on two datasets}
    \begin{tabular}{ccccccc}
    \toprule
    \multicolumn{1}{c}{\multirow{2}[2]{*}{Model}} & \multicolumn{1}{c}{\multirow{2}[2]{*}{Dimension}} & \multicolumn{1}{c}{\multirow{2}[2]{*}{Multi-hop}} & \multicolumn{2}{c}{Ciao} & \multicolumn{2}{c}{Epinions} \\
\cmidrule(lr){4-5} \cmidrule(lr){6-7}        &       &       & \multicolumn{1}{c}{Acc} & \multicolumn{1}{c}{F1} & \multicolumn{1}{c}{Acc} & \multicolumn{1}{c}{F1} \\
    \midrule
    \multicolumn{1}{c}{\multirow{6}[1]{*}{HGNN+}} & \multicolumn{1}{c}{\multirow{3}[1]{*}{64-32-16}} & 1  & 68.05 & 80.98 & 84.36 & 90.01 \\
          &       & 2     & \textbf{74.68} & \textbf{82.77} & \textbf{86.40} & \textbf{90.90} \\
          &       & 3     & 68.05 & 80.98 & 84.34 & 90.00 \\
          \cmidrule{2-7}  
          & \multicolumn{1}{c}{\multirow{3}[0]{*}{256-128-64}} & 1     & \textbf{82.28} & \textbf{88.00}    & \textbf{86.37} & \textbf{90.92} \\
          &       & 2     & 81.36 & 87.42 & 82.04 & 90.08 \\
          &       & 3     & 75.55 & 83.09 & 84.45 & 90.09 \\
          \cmidrule{1-7}
    \multicolumn{1}{c}{\multirow{6}[1]{*}{AHNTP}} & \multicolumn{1}{c}{\multirow{3}[0]{*}{64-32-16}} & 1     & 83.82 & 88.68 & 86.25 & 91.35 \\
          &       & 2     & \textbf{84.02} & \textbf{88.76} & \textbf{86.62} & \textbf{91.50} \\
          &       & 3     & 75.35 & 82.50  & 84.17 & 90.22 \\
          \cmidrule{2-7}
          & \multicolumn{1}{c}{\multirow{3}[1]{*}{256-128-64}} & 1     & \textbf{86.11}  & \textbf{90.11} & \textbf{89.78} & \textbf{92.94} \\
          &       & 2     & 81.21 & 87.11 & 85.50 & 90.37 \\
          &       & 3     & 68.94 & 81.25 & 85.68 & 90.26 \\
\bottomrule
\end{tabular}%
  \label{tab1}%
\end{table}

    \item \textbf{Analysis:} (1) AHNTP can have the best performance as it simultaneously considers the high-order social influence, an adaptive way to distinguish the different impacts of hyperedges, and contrastive learning. (2) The result of AHNTP outperforms AHNTP$_{\mathrm{nompr}}$, indicating the importance of high-order correlations in social influence for the prediction of trust. This result is consistent with the theory of Social Science, \ieno, users are more likely to trust people with a high social influence \cite{leskovec2013status, caverlee2008socialtrust, wang2015modeling, tang2016transfer}. This is because AHNTP can effectively identify users with high social influence based on their high-order correlation, which enhances the accuracy of representing node attributes and results in a more precise user representation. (3) AHNTP outperforms AHNTP$_{\mathrm{noatt}}$ because of the effectiveness of our adaptive GCN layers. By distinguishing the different impacts of hyperedges, the model can more focus on the relevant factors that have a significant impact on trust establishment. (4) AHNTP performs better than AHNTP$_{\mathrm{nocon}}$, reflecting the importance of supervised contrastive learning in trust prediction. This is because supervised contrastive learning can keep semantically similar and trusted users closer to each other while keeping dissimilar and distrusted users far away from each other in the embedding space, thereby enhancing the robustness and generalization of the model.
\end{itemize}


\subsection{Impact of Hyperparameters \textbf{(Q4)}}
\textbf{To answer Q4}, we investigate the influence of parameter tuning on multi-hop and hypergraph convolution layers, as well as the impact of the $\alpha$ and $t$ temperature parameters, on trust predictions.

\subsubsection{Multi-hop}
We use a multi-hop hypergroup to capture high-order neighbors reachable from each vertex. This approach can capture high-order correlations rather than simple pairwise correlations by incorporating groups of vertex information. Therefore, AHNTP can better mine the potential correlations within the data and yield a better representation. To find the influence of different hops on performance, we set the number of hops from 1 to 3. From TABLE VI, we can see that when the dimensions of the hypergraph convolutional layer are set to 256, 128, and 64, the performance ranking is 1-hop $>$ 2-hop $>$ 3-hop. However, when the dimensions are reduced to 64, 32, and 16, the performance ranking shifts to 2-hop $>$ 1-hop $>$ 3-hop. The performance declines with the increase of the hop, particularly at larger dimensions (256, 128, and 64). One possible reason is that the hypergraph structure accumulates extensive information from the higher-order neighbours that are far away from the node. Aggregating information from the neighbours with long distances may introduce noise labels and redundancies as not all the information from neighbours with long distances nodes is relevant. This aggregation may lead to signal dilution, resulting in a performance decline in AHNTP. When the dimensions of the hypergraph convolutional layers are reduced to smaller values (64, 32, and 16), which also reduces the noise, the results in 2-hop outperform those at 1-hop. This observation is consistent with the studies in HGNN+ \cite{gao2022hgnn}. These findings illustrate that considering multiple hops of neighbours in a hypergraph may not be able to improve the model performance due to the noise labels.

\subsubsection{Depth of Layers in AHNTP}
We vary the number of hypergraph convolution layers in AHNTP to investigate the optimal architecture. The layers are stacked from 1 to 5, and the experimental results are shown in Fig. \ref{image3} and Fig. \ref{image4}. AHNTP can have the best performance with 3 layers. After that, there is a notable decline in performance with the addition of each additional layer. This is because AHNTP aggregates high-order correlations from the neighbors with long distances. Therefore, with an increased number of layers, there can be over-smoothing issues.

\subsubsection{$\alpha$ Parameter}
\begin{figure}[ht]
  \centering
  \begin{minipage}[b]{0.8\linewidth}
    \includegraphics[width=\linewidth]{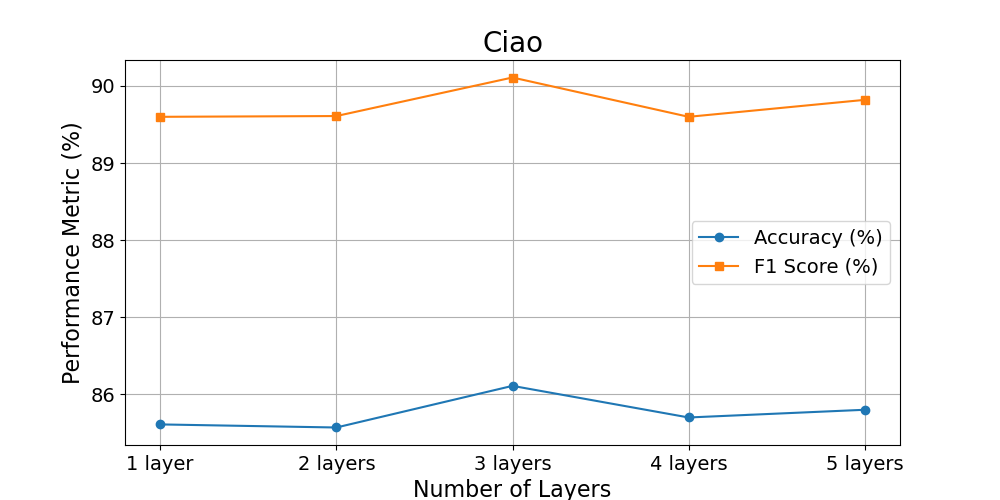}
    \caption{The Performance of AHNTP with Different Number of Layers on Ciao.}
    \label{image3}
  \end{minipage}
  \hfill
  \begin{minipage}[b]{0.8\linewidth}
    \includegraphics[width=\linewidth]{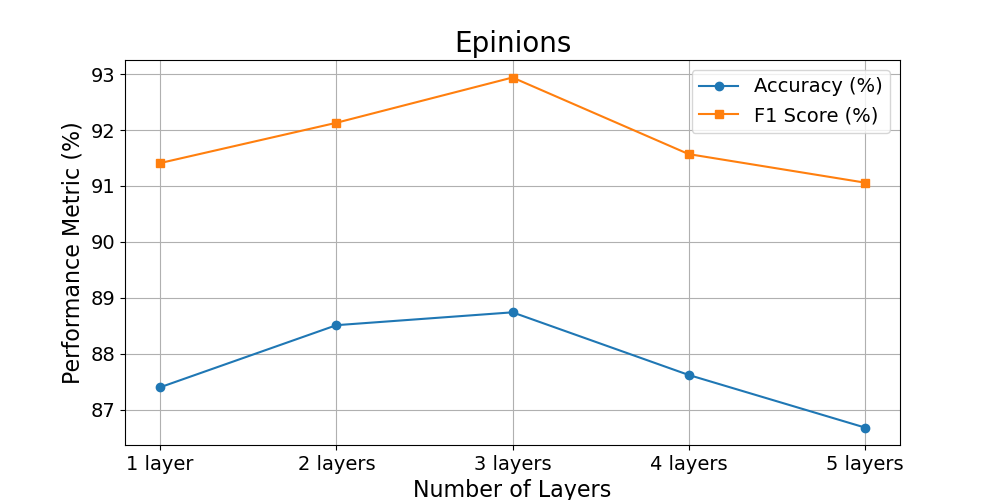}
    \caption{The Performance of AHNTP with Different Number of Layer on Epinions.}
    \label{image4}
  \end{minipage}
\hfill
  \begin{minipage}[b]{0.8\linewidth}
    \includegraphics[width=\linewidth]{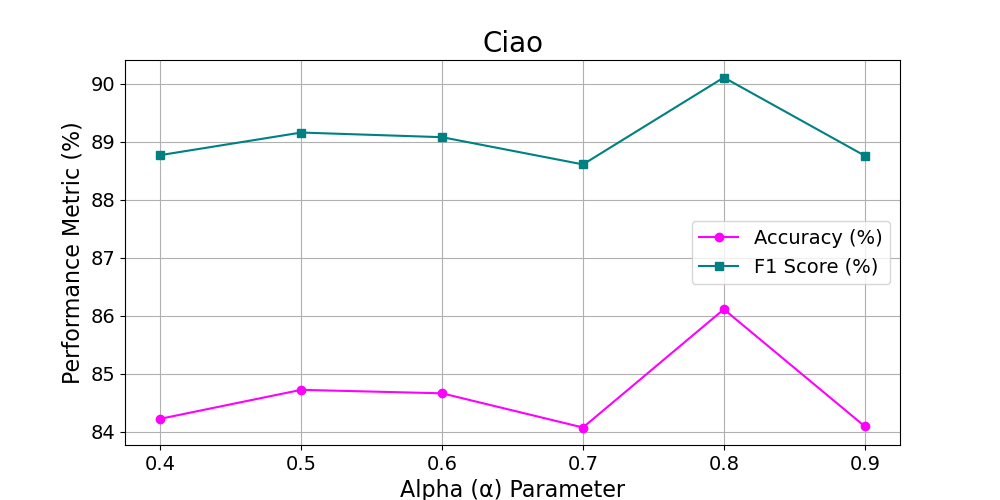}
    \caption{The Performance with Different $\alpha$ on Ciao.}
    \label{image5}
  \end{minipage}
\end{figure}

The parameter $\alpha$ is to control the equilibrium between edge-based correlation and motif-based high-order correlation. $\alpha = 0$ means the model does not include the high-order correlation based on motifs, whereas $\alpha = 1$ means that the model only considers the high-order correlation based on motifs. To investigate the impact of $\alpha$ on the performance of AHNTP, we set the value of $\alpha$ in the range of [0.4, 0.5, 0.6, 0.7, 0.8, 0.9] on both Ciao and Epinions datasets. The results are shown in Fig. \ref{image5} and Fig. \ref{image6}. Based on these results, we can observe that AHNTP can have the best performance when $\alpha$=0.8. 
This is because integrating both edge-based and motif-based high-order correlations can enrich the information for trust prediction.

\subsubsection{$t$ Temperature Parameter}
\begin{figure}[ht]
  \centering
  \begin{minipage}[b]{0.8\linewidth}
    \includegraphics[width=\linewidth]{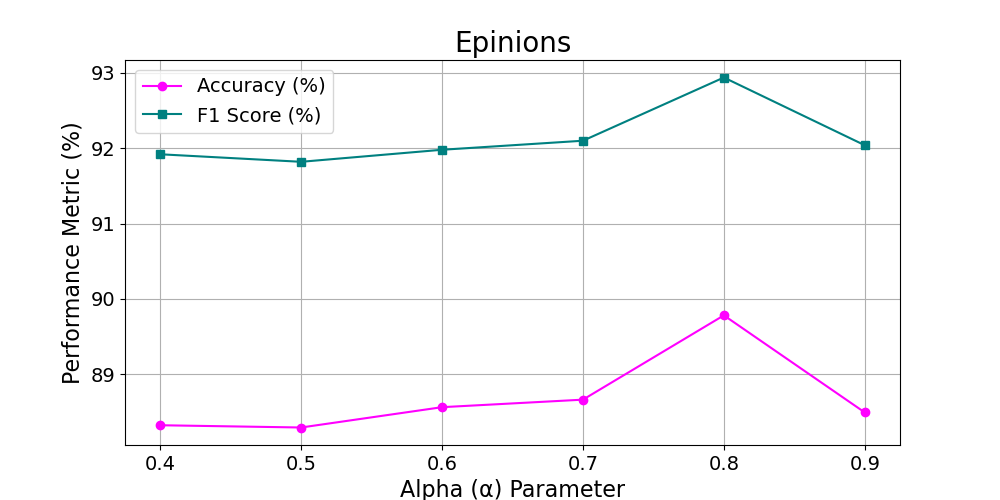}
    \caption{The Performance of Contrastive Learning with Different $t$ on Ciao.}
    \label{image6}
  \end{minipage}
  \hfill
  \begin{minipage}[b]{0.8\linewidth}
    \includegraphics[width=\linewidth]{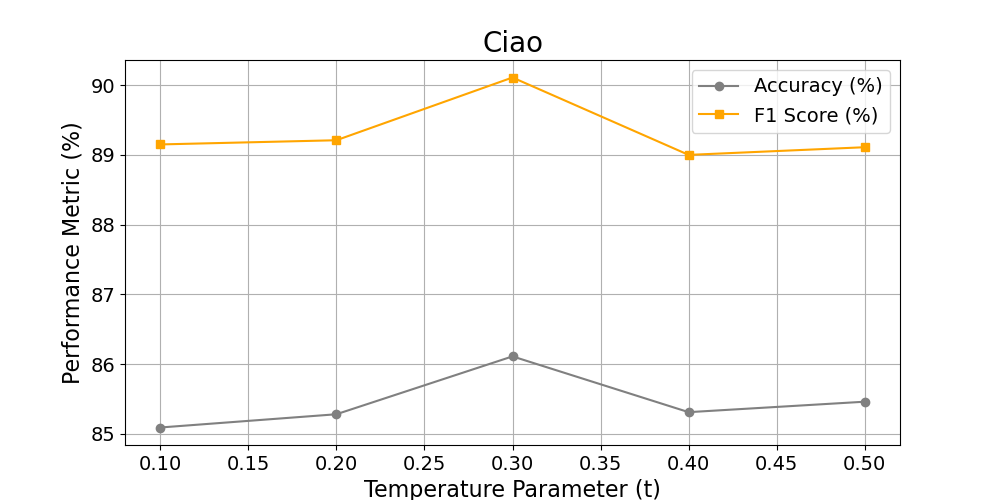}
    \caption{The Performance Metrics with Different $\alpha$ Parameter Values on Epinions.}
    \label{image7}
  \end{minipage}
  \hfill
  \begin{minipage}[b]{0.8\linewidth}
    \includegraphics[width=\linewidth]{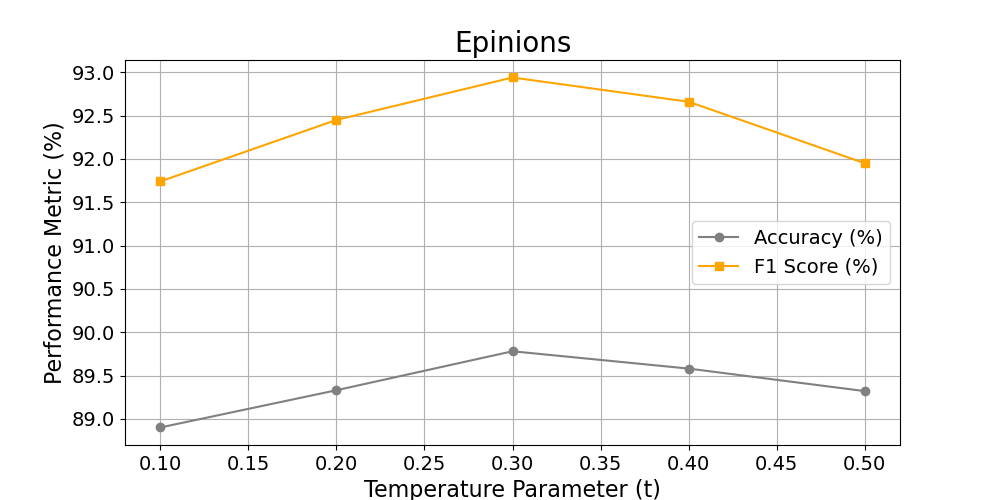}
    \caption{The Performance of Contrastive Learning with Different $t$ on Epinions.}
    \label{image8}
  \end{minipage}
\end{figure}

As discussed in Section IV, we employ the supervised contrastive learning method to control the distance between positive sample pairs and negative sample pairs. The temperature parameter, denoted as $t$, is an essential part of supervised contrastive learning. This parameter controls the distribution of the sample within the feature space. We set $t$ value as [0.1, 0.2, 0.3, 0.4, 0.5], to evaluate the performance of AHNTP. Fig. \ref{image7} and Fig. \ref{image8} show the results of the experiment on the Ciao and Epinions datasets, respectively. Based on these results, we can observe that the optimal value of $t$ is 0.3. This is because a large value of $t$ can lead to the over-smoothing problem on the probability distribution, which makes it difficult to distinguish the feature space.

\textbf{Summary.} The above experimental results can illustrate (1) The newly created hypergraph network in AHNTP is an effective method to capture high-order correlations in social networks. (2) The integration of social influence factors can incorporate more realistic information. (3) The adaptive hypergraph GCN layers effectively identify the dynamic influences of various factors. (4) The use of contrastive learning can contribute to generalization performance. In general, these features enable AHNTP to provide a comprehensive representation of complex social networks. Therefore, AHNTP outperforms the state-of-the-art methods in terms of the accuracy of trust prediction.

\section{Conclusion}
In this paper, we have proposed AHNTP, an innovative model for evaluating trust relationships in social networks. Initially, we applied the MPR algorithm to assign high-ranking scores to users with significant social influence. To address the issue of high-order correlations, we constructed node-level and structure-level attributes derived from the hypergraph. At the node level, attributes have included information on social influence and shared interests or activities of users. At the structural level, the attributes have captured the information of the user pairwise and multihop connections. Moreover, to identify the impact of the varying influence of hyperedges on users, we have developed the adaptive GCN layer to learn the node embedding. Furthermore, to improve the generalization and robustness performance of the model, we have introduced a supervised contrastive learning loss to optimize the model. Empirical evaluations of real-world datasets have demonstrated the efficiency and robustness of AHNTP. In our future work, we will devise a model for dynamic social networks that contain dynamic temporal information.

\bibliographystyle{IEEEtran}  
\bibliography{reference}     

\vspace{12pt}

\end{document}